\documentclass[sigconf,screen]{acmart}

\usepackage{xcolor}
\definecolor{deep-blue}{RGB}{0,0,200}
\definecolor{deep-red}{RGB}{200,0,0}
\definecolor{verde}{rgb}{0.25,0.5,0.35}
\definecolor{jpurple}{rgb}{0.5,0,0.35}
\definecolor{darkgreen}{rgb}{0.0, 0.2, 0.13}
\usepackage{listings}

\colorlet{BLUE}{blue}

\usepackage{multirow}
\usepackage{graphicx}
\usepackage{diagbox}
\usepackage{subfigure}
\usepackage{xspace}
\usepackage{algorithm}
\usepackage{bbding}
\usepackage{threeparttable}
\usepackage{amsmath}
\usepackage{adjustbox}
\usepackage{listings}
\usepackage[T1]{fontenc}
\usepackage{semantic}
\usepackage{newfloat}
\usepackage{enumitem}
\usepackage{url}
\usepackage{amsmath}
\usepackage{framed}
\usepackage{verbatim}
\usepackage[utf8]{inputenc}
\usepackage[T1]{fontenc}
\usepackage{microtype}
\usepackage{cleveref}
\usepackage{wrapfig}
\usepackage{mathtools}

\usepackage{pifont}
\usepackage{algorithm}
\usepackage{caption}
\usepackage[noEnd=true,commentColor=blue]{algpseudocodex}
\makeatletter
\@namedef{ver@lineno.sty}{9999/12/31}
\@namedef{opt@lineno.sty}{}
\makeatother
\usepackage[frozencache,cachedir=.]{minted}

\newcommand{\rvi}{RVV intrinsics}
\newcommand{\tool}{RVISmith}
\newcommand{\lmul}{\texttt{LMUL}}
\newcommand{\sew}{\texttt{SEW}}
\newcommand{\vlen}{\texttt{VLEN}}

\newcommand{\rv}{}

\begin{document}

\title{\tool{}: Fuzzing Compilers for RVV Intrinsics}


\author{Yibo He}
\orcid{0009-0002-1999-436X}
\affiliation{%
  \institution{Peking University}
  \department{Key Lab of HCST (PKU), MOE; SCS}
  \city{Beijing}
  \country{China}
}
\email{yibohe@pku.edu.cn}

\author{Cunjian Huang}
\orcid{0009-0008-4601-448X}
\affiliation{%
  \institution{DAMO Academy, Alibaba Group}
  \city{Hangzhou}
  \country{China}
}
\email{huangcunjian.huang@alibaba-inc.com}

\author{Xianmiao Qu}
\orcid{0009-0005-4990-1760}
\affiliation{%
  \institution{DAMO Academy, Alibaba Group}
  \city{Hangzhou}
  \country{China}
}
\email{xianmiao.qxm@alibaba-inc.com}

\author{Hongdeng Chen}
\orcid{0009-0004-4747-6345}
\affiliation{%
  \institution{DAMO Academy, Alibaba Group}
  \city{Hangzhou}
  \country{China}
}
\email{hongdeng.chd@alibaba-inc.com}

\author{Wei Yang}
\orcid{0000-0002-5338-7347}
\affiliation{%
  \institution{University of Texas at Dallas}
  \city{Richardson}
  \country{USA}
}
\email{wei.yang@utdallas.edu}

\author{Tao Xie}
\authornote{Corresponding author.}
\orcid{0000-0002-6731-216X}
\affiliation{%
  \institution{Peking University}
  \department{Key Lab of HCST (PKU), MOE; SCS}
  \city{Beijing}
  \country{China}
}
\email{taoxie@pku.edu.cn}

\begin{abstract}
Modern processors are equipped with single instruction multiple data (SIMD) instructions for fine-grained data parallelism.
Compiler auto-vectorization techniques that target SIMD instructions face performance limitations due to insufficient information available at compile time, requiring programmers to manually manipulate SIMD instructions.
SIMD intrinsics, a type of built-in function provided by modern compilers, enable programmers to manipulate SIMD instructions within high-level programming languages.
Bugs in compilers for SIMD intrinsics can introduce potential threats to software security, producing unintended calculation results, data loss, program crashes, etc.

To detect bugs in compilers for SIMD intrinsics, we propose RVISmith, a randomized fuzzer that generates well-defined C programs that include various invocation sequences of RVV (RISC-V Vector Extension) intrinsics.
We design RVISmith to achieve the following objectives: (i) achieving high intrinsic coverage, (ii) improving sequence variety, and (iii) without known undefined behaviors.
We implement RVISmith based on the ratified RVV intrinsic specification and evaluate our approach with three modern compilers: GCC, LLVM, and XuanTie.
Experimental results show that RVISmith achieves 11.5 times higher intrinsic coverage than the state-of-the-art fuzzer for RVV intrinsics.
By differential testing that compares results across different compilers, optimizations, and equivalent programs, we detect and report 13 previously unknown bugs of the three compilers under test to date.
Of these bugs, 10 are confirmed and another 3 are fixed by the compiler developers.
\end{abstract}



\maketitle
\footnotetext[1]{Accepted to ACM CCS 2025. This is the author's version for your personal use.}

\section{Introduction}~\label{sec:intro}
Modern processors typically support single instruction multiple data (SIMD) instructions, which perform operations on multiple data items in parallel.
To use SIMD instructions, programmers have three main approaches: (1) coding assembly instructions, being non-portable, error-prone, and extremely tedious, or (2) compiler auto-vectorization optimizations, or (3) manual vectorization by programming SIMD intrinsics~\cite{Neon, Intel-Intrinsics-Guide, RVV-intrinsic}.
Although persistent efforts have been made for automatic vectorization~\cite{vectorization-nips19, vectorization-CGO06, vectorization-pldi06, vectorization-pldi16}, compilers still face the inability to apply vectorization and non-optimal optimizations due to limited compile-time information~\cite{limit_autov1, vectorization-eval-taco19}.
SIMD intrinsics play a significant role in addressing the preceding limitation of automatic vectorization. 
As built-in functions inside modern compilers, SIMD intrinsics allow programmers to manipulate SIMD instructions like C functions in high-level programming languages.
Given the widespread reliance on SIMD intrinsics for performance-critical software, ensuring the correctness of their compilation is essential~\cite{compiler_bug_study}.

Despite this importance, detecting bugs of compilers for SIMD intrinsics receives little attention in existing research.
Previous compiler testing approaches, including both generation-based approaches (e.g., Csmith~\cite{Csmith}, YARPGen~\cite{YARPGen-oopsla20,YARPGen-pldi23}) and mutation-based approaches (e.g., equivalence modulo inputs (EMI) techniques~\cite{EMI,EMI2,EMI3}), are unable to generate any programs with SIMD intrinsics. 
Research work on SIMD-related topics such as automatic vectorization~\cite{vectorization-nips19, vectorization-CGO06, vectorization-pldi06, vectorization-pldi16, vectorization-eval-taco19, limit_autov1} and intrinsic evaluation~\cite{simdeval_16, simdeval_cgo18} traditionally focuses on performance but neglects the correctness of compilers for SIMD intrinsics.
Compilers for SIMD intrinsics are incorrectly presumed to be inherently robust and resistant to bugs, since most compilations from SIMD intrinsics to SIMD instructions are one-to-one translations.
To the best of our knowledge, no prior research work on compiler correctness for SIMD intrinsics has been identified in the research community.

To alleviate the preceding research gap, our work focuses on detecting compiler bugs for RVV (RISC-V Vector Extension) intrinsics.
RVV intrinsics are nascent and target the open RISC-V ISA, specifically requiring contributions from the open-source community.
RIF (\textbf{R}VV \textbf{I}ntrinsic \textbf{F}uzzing)~\cite{RIF} by SiFive is the only fuzzing tool available for \rvi{} that we are aware of, to the best of our knowledge. However, RIF faces inherent limitations, supporting only a restricted subset of intrinsics (less than 7\%) and a single operation per strip-mining loop (i.e., a loop that iterates over chunks or strips of data), due to the design to generate accurate calculation results as test oracles.
In practical use of \rvi{}, e.g., the deep-learning library OpenCV~\cite{OpenCV}, the combination of \rvi{} in a loop is much more complex than test cases generated by RIF.
A miscompilation bug of LLVM (\#\href{https://github.com/llvm/llvm-project/issues/106109}{106109}) related to a specific combination of \rvi{} has been reported and cannot be detected by RIF.

In this paper, we propose \tool{}, a randomized fuzzer that generates well-defined C programs that include various invocation sequences of \rvi{}. 
\tool{} addresses the following challenges.
\textbf{(1) Achieving high intrinsic coverage.} 
More than 120,000 intrinsics encoded with semantic information and corresponding vector types are defined in the RVV intrinsic document~\cite{RVV-intrinsic}, and a random combination of \rvi{} is error-prone.
We implement \tool{} with a novel technique to generate valid operation sequences in strip-mining loops, supporting more than 98\% \rvi{}.
\textbf{(2) Improving the sequence variety.}
We introduce vector register allocation and intrinsic scheduling to RVISmith to improve the sequence variety, including the variety of intrinsic combinations and data dependency inside intrinsic sequences.
Inspired by the general idea of EMI~\cite{EMI}, multiple constraints should be satisfied to ensure the semantic equivalence of different invocation sequences in the same random seed, enriching the test oracle of RVISmith.
\textbf{(3) Avoiding known undefined behaviors.}
Inheriting the unsafe tradition of C, undefined behaviors are ubiquitous in \rvi{}.
Undefined behaviors in \rvi{} exist in different forms from traditional C programs due to various operational semantics, memory-access vectorization, and implementation of inaccessible functions, leading to the failure of existing approaches to detect undefined behaviors, such as clang sanitizer~\cite{UndefinedBehaviorSanitizer} and STACK~\cite{ub_stack}.
Due to the lack of studying undefined behaviors in \rvi{}, we struggle with undefined behaviors in the development of \tool{}.
By systematically inspecting divergent execution cases and engaging with the RISC-V community, we apply multiple strategies to avoid undefined behaviors to both sequence generation and data generation.
We report the detected undefined behaviors of \rvi{} as a reference for future work, constituting an additional contribution.

\tool{} generates code with \rvi{} in four steps.
(1) Preprocessing and sequence selection. Given RVV-intrinsic definitions under test, \tool{} parses the definitions and stores relevant information with object-oriented data structures. 
\tool{} randomly selects a sequence of operation intrinsics based on a given ratio of \sew{} (i.e., selected element width in bits) and \lmul{} (i.e., length multiplier).
(2) Data-flow construction. We implement \tool{} with a random algorithm of register allocation that assigns variables to returned values and parameters in the selected operation intrinsics.
(3) Intrinsic scheduling. Intrinsic scheduling is to insert load intrinsics and store intrinsics into the selected operation intrinsics to construct a complete and valid invocation sequence of \rvi{}.
(4) Code generation. Code generated by \tool{} initializes element values in allocated memory, loads data from memory to vector-type variables, processes constants and data in vector-type variables, stores data from vector-type variables to memory, and prints values of well-defined elements to avoid undefined behaviors and detect bugs.
Additionally, \tool{} employs a differential testing framework that compares compilation and execution results across (1) different compilers in a single optimization, (2) a single compiler in different optimizations, and (3) equivalent programs generated by different intrinsic-scheduling algorithms.

To access the effectiveness of \tool{}, we evaluate \tool{} with three modern compilers: GCC, LLVM, and XuanTie. 
Our experiments show that \tool{} achieves 11.5 times intrinsic coverage higher than RIF, the state-of-the-art fuzzing tool for \rvi{}.
\tool{} detects 13 previously unknown bugs in the three compilers under test.
Among these bugs, 10 are confirmed and another 3 are fixed by the compiler developers.
More than 20,000 \rvi{} are affected by these bugs.
Most of the bugs are miscompilations that are difficult to detect and harmful to software security, leading to unintended calculation results, data loss, and emulator crashes without any compiler warning messages.
Moreover, numerous cases generated by \tool{} are found to lead to incorrect results compiled by historical versions of GCC and LLVM (but are correct by the latest version).
This finding shows that \tool{} also detects many known compiler bugs.
The exact number of known bugs is not reported due to the extensive labor of classification.

In summary, we make the following main contributions:
\begin{itemize}
\item \textbf{Implementation of fuzzer.} We propose \tool{} that randomly generates well-defined programs with \rvi{}. \tool{} is the first tool that can generate complex combinations of \rvi{} in strip-mining loops, supporting almost all \rvi{}.
\item \textbf{Detection of real-world bugs.} We detect 13 previously unknown bugs of GCC, LLVM, and XuanTie, improving the reliability and security of mainstream compilers.
\item \textbf{Empirical study of bugs.} We evaluate \tool{} with multiple historical versions of GCC and LLVM. We report empirical evaluation results showing that compiler bugs related to \rvi{} exist widely in versions of GCC and LLVM.
\item \textbf{Undefined-behavior report.} We make the first report of undefined behaviors caused by \rvi{} and how we deal with them in \tool{}. 
These undefined behaviors lead to typical unsafe issues such as the use of uninitialized variables, and out-of-bound writes.
Any future work based on \rvi{} can refer to our report.
\end{itemize}

The implementation of \tool{} based on the ratified RVV intrinsic document in version 1.0~\cite{RVV-intrinsic} is available at \url{https://github.com/yibo2000/RVISmith}, and our artifacts are available at \url{https://zenodo.org/records/15548270}.
\tool{} represents a fresh start for both academia and industry in detecting potential compiler bugs related to built-in functions, especially SIMD intrinsics.

\section{Background}\label{sec:background}
This section presents some domain knowledge about \rvi{}.
Two issues are discussed in this section: (1) reasons why modern compilers are equipped with SIMD intrinsics, and (2) a brief introduction to \rvi{}.

\begin{figure*}[t]
    \centering
    \begin{minipage}{\textwidth}
        \begin{minipage}[b]{0.25\textwidth}
            \centering
            \begin{minted}[fontsize=\small]{c}
for(int i=0; i<length; i++){
    c[i] = a[i] + b[i];
}




            \end{minted}
            \caption*{(a) Scalar code}
    \end{minipage}
        \hfill
    \begin{minipage}[b]{0.45\textwidth}
            \centering
            \begin{minted}[fontsize=\small]{c}
for (int avl=length; avl>0; avl-=vl,a+=vl,b+=vl,c+=vl) {
    vl = __riscv_vsetvl_e32m2(avl);
    vfloat32m2_t va = __riscv_vle32_v_f32m2(a, vl);
    vfloat32m2_t vb = __riscv_vle32_v_f32m2(b, vl);
    vfloat32m2_t vc = __riscv_vfadd_vv_f32m2(va, vb, vl);
    __riscv_vse32_v_f32m2(c, vc, vl);
}
            \end{minted}
            \caption*{(b) SIMD intrinsic code}
        \end{minipage}
        \hfill
        \begin{minipage}[b]{0.25\textwidth}
            \centering
            \begin{minted}[fontsize=\small]{asm}
vsetvli zero,a5,e32,m2,ta,ma
vle32.v v2,0(a0)
vle32.v v4,0(a1)
vfadd.vv    v2,v2,v4
vse32.v v2,0(a2)


            \end{minted}
            \caption*{(c) SIMD assembly instructions}
        \end{minipage}
    \end{minipage}
    \caption{Three implementations for adding two 32-bit floating-point arrays. (b) and (c) are for RISC-V V extension.}
    \label{fig:example_code}
\Description[<short description>]{<long description>}
\end{figure*}

\subsection{Why SIMD Intrinsics?}\label{sec2.1}
Domains such as machine learning, image processing, and cloud computing have caused an increase in the significance and sophistication of single instruction multiple data (SIMD) instructions.
The key idea of SIMD is, in a single instruction, to calculate multiple data elements simultaneously (e.g., (b) and (c) in Figure~\ref{fig:example_code}) rather than data elements one by one (e.g., (a) in Figure~\ref{fig:example_code}). 
Compared to the basic single instruction with single data, SIMD instructions can greatly improve data processing performance.
Most modern processors support SIMD instructions.

There are three main approaches for programmers to use SIMD instructions.
(1) Embedded assembly or assembly instructions (e.g., Figure~\ref{fig:example_code}(c)). 
Programmers can code an assembly snippet with SIMD instructions, and then embed the assembly snippet to code in high-level programming languages or directly obtain an executable file from the assembly snippet.
(2) Compiler optimizations by automatic vectorization (e.g., translation from Figure~\ref{fig:example_code}(a) to Figure~\ref{fig:example_code}(c)). 
Modern compilers typically are implemented with auto-vectorization optimizations that generate SIMD instructions from scalar code, including loop-level vectorization~\cite{vectorization-pldi06,vectorization-pldi16} and superword-level parallelism (SLP)~\cite{vectorization-nips19,vectorization-asplos21}.
(3) Manual vectorization by SIMD intrinsics (e.g., Figure~\ref{fig:example_code}(b)). 
SIMD intrinsics are designed to encapsulate SIMD instructions, allowing programmers to manipulate SIMD instructions like C functions.
SIMD intrinsics are typically built-in functions, with functionality implemented by compilers.
Compilers release programmers from tedious procedures for using SIMD instructions such as register allocation, and setting control and status registers.

The preceding three approaches for using SIMD instructions each have their own applicable scenarios.
Assembly instructions can be used for extremely fine-grained optimization even at clock-cycle level.
However, coding assembly instructions is extremely error-prone and tedious, and assembly instructions lack platform portability, resulting in very limited applicable scenarios of coding SIMD assembly instructions for data parallelism.
Compiler optimizations by automatic vectorization can improve the performance of data processing.
However, the ability of compilers for auto-vectorization depends on the compiler's capability at compile time to analyze a program for precise information~\cite{RefinedInputDegradedOutput}, and the actual performance after auto-vectorization optimizations is far from the architectural peak performance due to various obstacles, non-optimal optimizations, and inability to obtain information at compile time~\cite{limit_autov1, vectorization-eval-taco19}.
Given the preceding limitations of coding assembly and compiler optimizations by automatic vectorization, SIMD intrinsics are widely used in modern processors~\cite{Intel-Intrinsics-Guide, Neon, RVV-intrinsic}.

\subsection{RVV Intrinsics}\label{sec2.2}
In this section, we provide a brief introduction to \rvi{}. 
The domain knowledge of \rvi{} is integral to the design of \tool{}, as \rvi{} are the primary subjects under test.
For a complete introduction, please refer to the ratified RVV intrinsic document~\cite{RVV-intrinsic}.

\begin{figure}
  \begin{center}
    \includegraphics[width=0.36\textwidth]{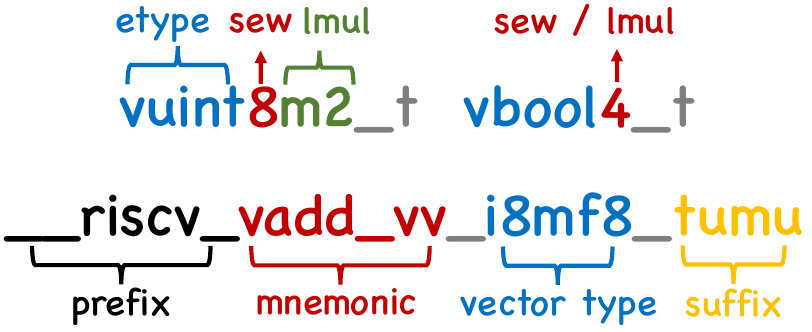}
  \end{center}
  \vspace{-0.2cm}
  \caption{Diagram of vector type naming and intrinsic naming for \rvi{}.} 
  \label{fig:datatype}
\Description[<short description>]{<long description>}
\end{figure}

\textbf{Type system.} 
SIMD instructions are typically used together with a group of vector registers that have larger lengths than regular registers for holding multiple elements during vectorization. 
The number of bits in a single vector register in RVV is marked as \vlen{}, which is an implementation-defined constant parameter.
\sew{} (Selected Element Width) is dynamic to determine the size of elements in bits being processed.
By default, a vector register is viewed as being divided into \vlen{}/\sew{} elements by default.
RVV also supports the length multiplier (\lmul{}), which allows us to process a single vector register ($\lmul{}=1$), multiple vector registers as a vector group ($\lmul{}>1$), and a fraction of a vector register ($\lmul{}<1$).

\rvi{} are equipped with an extended type system, encoding information such as \sew{} and \lmul{} into vector types shown in Figure~\ref{fig:datatype}.
For example, \texttt{vuint8m2\_t} represents that elements of this type use two vector registers as a register group, and each element in the register group is an 8-bit unsigned integer.
An exception of type naming in the type system of \rvi{} is the bool vector type.
As we all know, a bool element requires only one bit, and for this reason, a bool vector type uses only one vector register to optimize resource usage.
The first $\vlen{} \times \lmul{} \div \sew{}$ (which is less than \vlen{}) bits of the bool vector register represent the valid bool elements, and the number in a bool vector type represents the ratio of \sew{}/\lmul{} as shown in Figure~\ref{fig:datatype}.

\textbf{Intrinsic naming scheme.} 
The names of RVV intrinsics can generally be divided into four parts as shown in Figure~\ref{fig:datatype}.
(1) Prefix. In the ratified version of RVV intrinsic document~\cite{RVV-intrinsic}, all intrinsic names have an identical prefix \texttt{\_\_riscv\_} to avoid potential naming conflicts.
(2) Mnemonic. A mnemonic is the RVV instruction name after replacing the dots with underscores. For example, the \texttt{\_\_riscv\_vadd\_vv\_i8mf8\_tumu} intrinsic in Figure~\ref{fig:datatype} uses a mnemonic from the \texttt{vadd.vv} instruction, which means adding two signed integer vectors.
(3) Vector type. 
RVV intrinsic names are explicitly encoded with the main vector type used in the calculation, e.g., \texttt{i8mf8} in the \texttt{\_\_riscv\_vadd\_vv\_i8mf8\_tumu} intrinsic in Figure~\ref{fig:datatype}.
For most \rvi{}, the main vector types are unique.
A small portion of \rvi{} are encoded with more than one vector type such as \texttt{\_\_riscv\_vreinterpret\_v\_i8mf8\_u8mf8} to avoid naming conflicts.
(4) Suffix. A suffix in an intrinsic name defines whether the intrinsic is a masking operation, how to deal with masked-off elements, and how to deal with tail elements. 
Masked-off elements are those elements that do not need to be operated, and tail elements are elements in unused positions of a vector register.
There are two options to deal with masked-off elements and tail elements in RVV: undisturbed (i.e., keeping original values) and agnostic (i.e., unknown values).

There are four types of RVV intrinsic names to avoid always writing long names during using \rvi{}.
(1) Explicit (non-overloaded) intrinsics. 
\rvi{} in this type do not have policy suffixes that define how to deal with masked-off elements and tail elements.
The default option is tail agnostic and masked-off agnostic.
(2) Explicit (non-overloaded) intrinsics, policy variants.
\rvi{} in this type have all four parts in names shown in Figure~\ref{fig:datatype}.
(3) Implicit (overloaded) intrinsics.
Most intrinsics of this type are not encoded with vector types or policy suffixes.
(4) Implicit (overloaded) intrinsics, policy variants.
Most of intrinsics in this type are not encoded with vector types.

\textbf{Control and Status Registers (CSRs).}
Multiple CSRs exist in the RVV programmer model, including seven unprivileged CSRs added by the V extension, CSRs in the base ISA and CSRs in other extensions.
CSRs are not directly controlled by intrinsic programmers.
Intrinsic programmers can specify or get the status of partial CSRs by calling related intrinsics and setting corresponding arguments.
Other CSRs (such as \texttt{vstart}) that are not exposed to the intrinsic level are excluded from being controlled at the intrinsic level.
Compilers for \rvi{} are responsible for generating correct instructions that read and write CSRs.

We provide a brief introduction of intrinsics and parameters related to CSRs.
The parameter ``\texttt{unsigned int frm}'' specifies the floating rounding mode CSR (\texttt{frm}) for partial floating-point intrinsics.
The parameter ``\texttt{unsigned int vxrm}'' specifies the fixed-point rounding mode CSR (\texttt{vxrm}) for most fixed-point intrinsics.
The \texttt{\_\_riscv\_vlenb} intrinsic returns the value inside the read-only CSR \texttt{vlenb}.

\texttt{vtype} and \texttt{vl} CSRs are special, which do not require explicit values when intrinsics are programmed.
The \texttt{vtype} CSR provides how to interpret the contents of vector registers, including \sew{}, \lmul{}, whether masked-off elements are agnostic, whether tail elements are agnostic, etc.
The \texttt{vl} CSR provides an unsigned integer specifying the number of elements to be updated by a vector instruction.
Compilers are responsible for controlling the status of \texttt{vtype} and \texttt{vl} CSRs during translating each RVV intrinsic.
At the intrinsic level, the \texttt{vtype} CSR is specified by intrinsic names as discussed earlier, and the \texttt{vl} CSR is mostly related to the ``\texttt{size\_t vl}'' parameter that exists in most \rvi{}.
Under the programming specification of \rvi{}, programmers obtain a size\_t value by appropriate \texttt{vsetvl} or \texttt{vsetvlmax} intrinsics, and then use the value as the argument for the ``\texttt{size\_t vl}'' parameter, which specifies the number of elements to be updated in each iteration.
Formulas of \texttt{vsetvl} and \texttt{vsetvlmax} intrinsics are shown as follows:
\begin{align*}
vsetvl(avl) &= \min(avl, vlmax) & vsetvlmax() &= vlmax
\end{align*}
\[vlmax = \vlen{} \times \lmul{} \div \sew{}\]
In the formula, \texttt{avl} is the application vector length, which represents the length of the remaining vector to be processed in the program. 
\texttt{VLMAX} is the maximum number of elements in a vector that one RVV intrinsic can process, given \sew{} and \lmul{}.
From this formula, we can know that the maximum number of elements in each intrinsic's iteration is equal if the ratio of corresponding \sew{} / \lmul{} is equal.
\section{\tool{}}\label{sec:approach}

\begin{figure*}[t]
\includegraphics[width=0.9\textwidth]{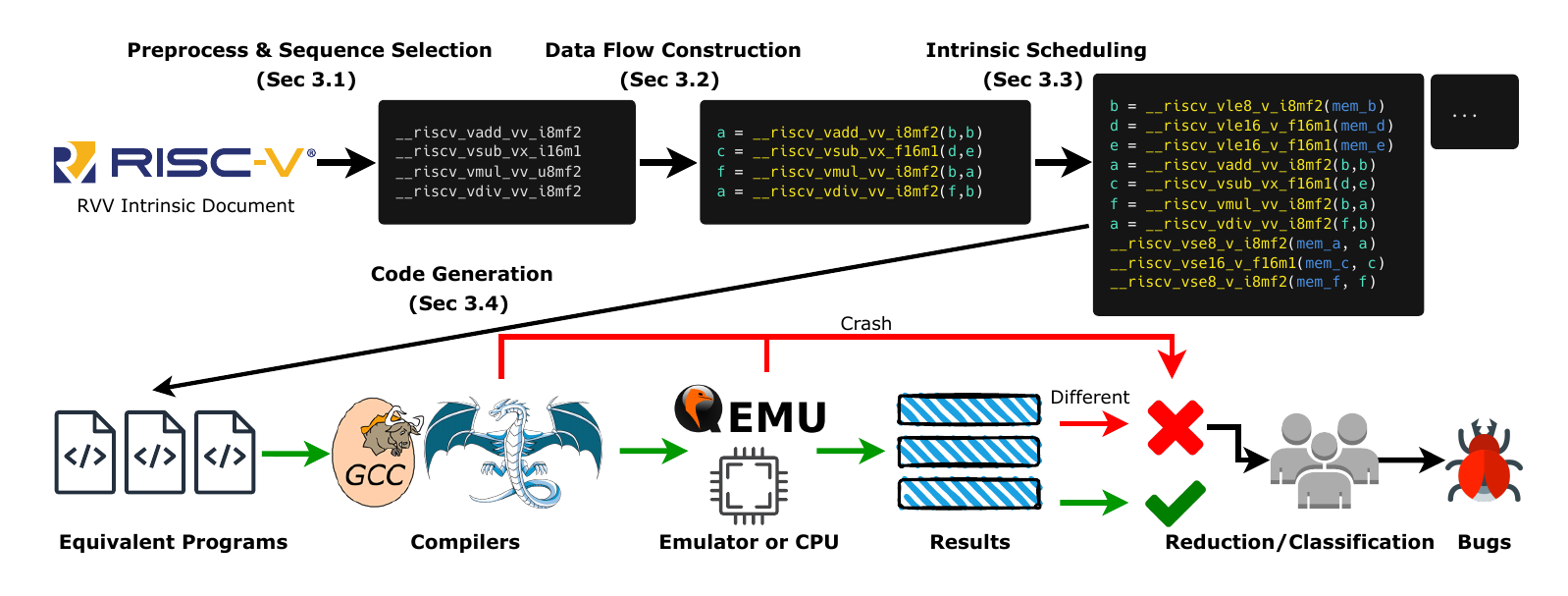}
\centering
\vspace{-0.2cm}
\caption{Overview of \tool{}.}
\label{fig:overview}
\Description[<short description>]{<long description>}
\end{figure*}

Figure~\ref{fig:overview} shows an overview of \tool{}'s workflow.
\rv{Before delving into the technical details, we briefly introduce (1) how \tool{} generates well-defined programs, and (2) how we use the generated programs to detect compiler bugs by differential testing.}

\rv{\textbf{Generation of well-defined programs.}}
\tool{} parses the document of \rvi{} and selects a ratio-based sequence of \rvi{} (Section~\ref{sec:3.1}).
This step ensures that a representative and diverse mix of intrinsics are covered during testing, reflecting realistic usage patterns.
Then, \tool{} constructs data dependencies inside the sequence based on a randomized algorithm of vector-register allocation (Section~\ref{sec:3.2}).
This step introduces realistic data dependency chains.
Next, \tool{} performs intrinsic scheduling to insert load intrinsics and store intrinsics and constructs multiple equivalent sequences by different scheduling algorithms (Section~\ref{sec:3.3}).
These variants enable broader coverage and robustness in differential testing.
Finally, \tool{} generates complete programs, adding code snippets that initialize allocated memory and pointers, update loop variables, and print non-agnostic elements for differential testing (Section~\ref{sec:3.4}).
We also discuss undefined behaviors of \rvi{} found by us and how \tool{} avoids these undefined behaviors (Section~\ref{sec:3.5}).

\rv{\textbf{Differential testing.}
Three differential-testing strategies are used in our work after generating programs with \rvi{}, comparing compilation and execution results from (1) different compilers in the same optimization, (2) the same compiler in different optimizations, and (3) equivalent programs (that are generated by different intrinsic-scheduling algorithms) compiled by the same compiler in the same optimization.
Any compiler crashes, runtime crashes, and different execution results indicate detected bug cases (mainly compiler bugs).
We manually minimize and classify the detected bug cases by delta debugging, observing error behaviors, and communicating with developers.}

The experiments in this work focus on only compiler fuzzing.
In theory, \tool{} can also be used to test emulators and CPUs for RVV instructions by differential testing on different hardware, but this topic is beyond the scope of our paper.

\subsection{Preprocessing and Sequence Selection}\label{sec:3.1}
The preprocessing procedure consists of two key aspects given the document of \rvi{} under test.
(1) RVISmith filters out irrelevant text and parses the given RVV-intrinsic definitions in the document.
\tool{} allows users to test any subset of all \rvi{}, as some compilers support only partial \rvi{} and some processors support only partial RVV instructions.
For example, \rvi{} involving 64-bit elements are not supported by 32-bit processors.
(2) \tool{} divides \rvi{} under test into four categories according to intrinsic functionality by a static analysis of definitions.
As introduced in Section~\ref{sec2.2}, the intrinsic name portion in an RVV-intrinsic definition is encoded with how the intrinsic is used (e.g., corresponding assembly instruction, return type, parameter list, policy), and \tool{} constructs objects to store this information by our object-oriented data structures.
The four categories are as follows:

\begin{itemize}[leftmargin=*]
    \item Load intrinsics: reading data from memory to vector-type variables. Load intrinsics can be recognized by matching mnemonic with RVV load instructions such as \texttt{vle\{i\}.v}.
    \item Store intrinsics: writing data from vector-type variables to memory. Store intrinsics are void return functions with the common prefix \texttt{\_\_riscv\_vs} in the design of \rvi{}.
    \item Ignored intrinsics: \texttt{vsetvl} intrinsics, \texttt{vsetvlmax} intrinsics, and unsupported intrinsics (mainly fault-only-first load intrinsics).
    \item Operation intrinsics: processing constants and vector-type variables. All intrinsics outside the preceding three categories are classified as operation intrinsics by \tool{}.
\end{itemize}

After preprocessing, \tool{} randomly selects a ratio-aligned sequence (Definition~\ref{definition-2}) from operation intrinsics under test.
This sequence represents the computational operations in a strip-mining loop \rv{(i.e., a loop that iterates over chunks of data)} in the generated code.
The current version of \tool{} generates only one strip-mining loop with multiple \rvi{} in each program.
Sequence selection in \tool{} is based on a specified ratio of \sew{} / \lmul{} \rv{(as discussed in Section~\ref{sec2.2}, \sew{} is the size in bits of elements that are being processed, and \lmul{} is the length multiplier)}.
We provide the definition of ratio-aligned intrinsics and the definition of ratio-aligned intrinsic sequences as follows:
\begin{definition}[Ratio-aligned intrinsic]\label{definition-1}
An intrinsic with at least one vector type is ratio-aligned, if and only if all vector types in the intrinsic (including return type and parameter type) share the same ratio of \sew{} / \lmul{}, and the ratio is the common ratio of this intrinsic.
\end{definition}

\begin{definition}[Ratio-aligned intrinsic sequence]\label{definition-2}
A sequence of \rvi{} is ratio-aligned, if and only if 
(1) all ratio-aligned intrinsics in the sequence share the same common ratio, and 
(2) all intrinsics in the sequence that are not ratio-aligned include at least one vector type with the same ratio as the common ratio of ratio-aligned intrinsics in the sequence.
\end{definition}

The reason that sequences selected by \tool{} should be ratio-aligned is as follows. 
\tool{} seeks to generate VLA-Style (Vector Length Agnostic Style) programs that are portable under different \vlen{} settings, \rv{i.e., the number of bits (in a single vector register) that are decided by processors.}
A ratio-aligned intrinsic sequence ensures that all elements of an input vector can be processed in the sequence's strip-mining loop without skipped elements, as \texttt{vsetvl} and \texttt{vsetvlmax} intrinsics return the identical unsigned integer (determining how many elements are processed per iteration) given the same ratio of \sew{} / \lmul{} as discussed in Section~\ref{sec2.2}.
If a sequence of \rvi{} is not ratio-aligned and uses \texttt{vsetvl} intrinsics as arguments for \texttt{vl} parameters and to update loop variables, elements are either processed repeatedly or skipped, and the number of skipped elements depends on the return values of \texttt{vsetvl} intrinsics, resulting in a violation of the VLA-Style.
For example, as the code in Figure~\ref{fig:non-ratio-aligned-seq}, 1/2 \texttt{vl} (when $avl > vlmax$) of elements are skipped by \texttt{f32m2} intrinsics as the \texttt{vl} is for the preceding \texttt{e32m4} intrinsics.

\begin{figure}[t]
\begin{minipage}[t]{\linewidth}
\begin{minted}[fontsize=\small]{c}
    vl = __riscv_vsetvl_e32m4(avl);
    /* some e32m4 (ratio=8) intrinsics */
    vfloat32m2_t va = __riscv_vle32_v_f32m2(a, vl);
    vfloat32m2_t vb = __riscv_vle32_v_f32m2(b, vl);
    vfloat32m2_t vc = __riscv_vfadd_vv_f32m2(va, vb, vl);
    __riscv_vse32_v_f32m2(c, vc, vl); //ratio=16
    a += vl; b += vl; c += vl;
    avl -= vl;
\end{minted}
\end{minipage}%

\centering
\caption{An example of non-ratio-aligned RVI sequence.}
\label{fig:non-ratio-aligned-seq}
\Description[<short description>]{<long description>}
\end{figure}

To select a ratio-aligned intrinsic sequence, \tool{} extracts the ratio of \texttt{SEW} / \texttt{LMUL} from a user-specified vector type, filters out all intrinsics that can make up a ratio-aligned intrinsic sequence given the ratio, and randomly selects \textit{n} (specified by users) operation intrinsics from the filtered intrinsics.
There are two types of exceptions.
(1) Type conversion intrinsics. 
Type conversion intrinsics always lead to undefined behaviors.
We discuss how to deal with these intrinsics in Section~\ref{sec:3.5}.
(2) Reduction intrinsics.
During using reduction intrinsics, only the \texttt{vs2} parameter (i.e., the second parameter under operation except masking) is iterated in the loop to reduce the vector dimensionality, such as intrinsics for the \texttt{vredor} instruction ($vd[0] = or( vs1[0] , vs2[*]$, where [*] denotes all active elements).
For reduction intrinsics, \tool{} selects only those whose ratios of types of the \texttt{vs2} parameter are the same as the given common ratio.

\subsection{Data-Flow Construction}\label{sec:3.2}

We implement \tool{} with a randomized algorithm of vector-register allocation, which assigns variables to parameters and returned values in the sequence of ratio-aligned operation intrinsics selected by \tool{}, making programs generated by \tool{} (1) follow the use-define chain convention, and (2) cover all four data dependency scenarios (read-read, read-write, write-read, and write-write).
Given that the main types of \rvi{} are vector types, \tool{} focuses on the allocation of variables in vector types.
For scalars and CSRs in parameters, \tool{} randomly generates corresponding constants during code generation.

\begin{algorithm}[t]
\caption{Vector-Register Allocation}\label{alg:register-allocation}
\begin{algorithmic}[1]

\Procedure{VRegisterAllocation}{\textit{Operation intrinsics $I$}}
\State VregTable $\gets$ \{ \}
\ForAll{$op \in I$}

\ForAll{$p \in op.vector\_parameters$}
\If{CoinFlip() or VregTable[$p.type$] is [ ]}
\LComment{A load intrinsic for vreg\_mem is required for intrinsic scheduling.}
\State $p.vreg \gets \textbf{new}$ vreg\_mem
\State VregTable[$p.type$].append($p.vreg$)
\Else
\State $p.vreg \gets$ randomSelect(VregTable[$p.type$])
\EndIf
\EndFor

\For{ $ret \gets op.vector\_ret$ }
\If{CoinFlip() or VregTable[$ret.type$] is [ ]}
\LComment{No load intrinsic for vreg is required for intrinsic scheduling.}
\State $ret.vreg \gets \textbf{new}$ vreg
\State VregTable[$ret.type$].append($ret.vreg$)
\Else
\State $ret.vreg \gets $randomSelect(VregTable[$ret.type$])
\EndIf
\EndFor
\EndFor
\State \Return $I$
\EndProcedure

\end{algorithmic}
\end{algorithm}

Given a sequence of selected operation intrinsics, vector-register allocation in \tool{} works as Algorithm~\ref{alg:register-allocation}.
\tool{} maintains a key-value table of vector registers during register allocation.
Keys of the vector-register table are strings that represent vector types, and values of the vector-register table are arrays of strings that represent active variables in the corresponding vector types.
Whenever a new register $R$ in type $T$ is allocated, the vector-register table appends $R$ to the array of $T$.
The function \texttt{CoinFlip()} randomly returns either \texttt{true} or \texttt{false}, which determines whether \tool{} allocates a new register or uses a currently active register.
Vector registers newly allocated for parameters of operation intrinsics have a common suffix \texttt{\_mem} (Line 7), which means that a load intrinsic for this register is required for the following intrinsic scheduling.

\subsection{Intrinsic Scheduling}\label{sec:3.3}
Intrinsic scheduling is to obtain a complete invocation sequence of \rvi{} by inserting load intrinsics and store intrinsics into selected operation intrinsics.
For each vector variable in parameters of operation intrinsics, if this variable occurs for the first time, a load intrinsic should be inserted before the operation intrinsic.
For each vector variable assigned by return values of operation intrinsics, a store intrinsic should be inserted after the operation intrinsic.

We model the problem of intrinsic scheduling as follows.
Given a sequence of operation intrinsics $I$ after data flow construction, for each intrinsic \rv{$I[i] \in I$}, there is an array of prefix intrinsics \rv{$P[i]$} (mainly load intrinsics) that should be called before \rv{$I[i]$}, and an array of suffix intrinsics \rv{$S[i]$} (mainly store intrinsics) that should be called after \rv{$I[i]$}. 
Let \rv{\texttt{N}} be length of the sequence of operation intrinsics \rv{(i.e., common size)}, the intrinsic scheduling should satisfy the following constraints for syntactic correctness:
\rv{\begin{itemize}[leftmargin=*]
    \item For each $i \in range(0, N)$, all intrinsics in $P[i]$ should be executed before $I[i]$, and all intrinsics in $S[i]$ should be executed after $I[i]$.
    \item For any $i, j \in range(0, N)$, if $i < j$, $I[i]$ should be executed before $I[j]$.
    \item For any $x \in range(0, N)$, for any $i, j \in range(0, P[x].size)$, if $i < j$, $P[x][i]$ should be executed before $P[x][j]$.
    \item For any $x \in range(0, N)$, for any $i, j \in range(0, S[x].size)$, if $i < j$, $S[x][i]$ should be executed before $S[x][j]$.
\end{itemize}}

To satisfy the preceding constraints, we implement \tool{} with three intrinsic-scheduling algorithms. 
We provide the intrinsic-scheduling algorithms in \rv{Algorithms~\ref{alg:scheduling-allin},~\ref{alg:scheduling-unit}, and~\ref{alg:scheduling-random}}.
(1) \textbf{All-in intrinsic scheduling} (in Algorithms~\ref{alg:scheduling-allin}). 
All prefix intrinsics are called at the beginning of the sequence (Lines 3-5). 
\rv{Operation intrinsics are called between prefix intrinsics and suffix intrinsics (Lines 6-7).}
All suffix intrinsics are called at the end of the sequence (Lines 8-10).
(2) \textbf{Unit intrinsic scheduling} (in Algorithms~\ref{alg:scheduling-unit}). 
Prefix intrinsics and suffix intrinsics are called immediately before/after the corresponding operation intrinsic \rv{(Lines 3-8)}. 
(3) \textbf{Random intrinsic scheduling} (in Algorithms~\ref{alg:scheduling-random}). 
Intrinsics can be called at any location that satisfies the preceding constraints \rv{(Lines 4-11)}. 
The function \texttt{randomInsert} randomly inserts a value between the begin pointer and the end pointer and returns the pointer to the inserted value.

\begin{figure}[t]
\includegraphics[width=0.475\textwidth]{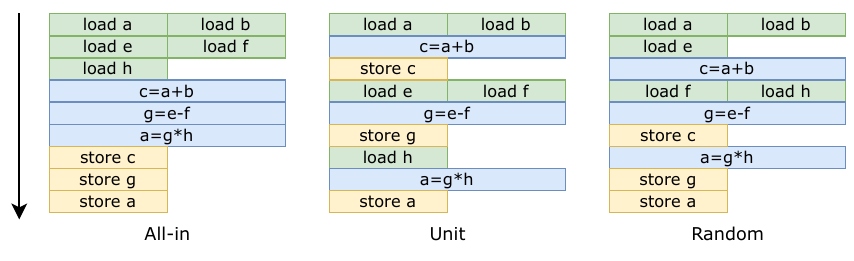}
\centering
\caption{\rv{An example of invocation sequences after intrinsic scheduling.}}
\label{fig:scheduling}
\Description[<short description>]{<long description>}
\end{figure}

\rv{Figure~\ref{fig:scheduling} shows an example of invocation sequences after intrinsic scheduling for a more precise presentation.
The all-in algorithm and the unit algorithm represent extreme cases: prefix/suffix intrinsics are called at the beginning/end, or adjacent to the corresponding operation intrinsics.
The random algorithm represents general cases: all well-defined invocation sequences can occur.}

\textbf{Equivalent programs.} Programs generated by \tool{} with different scheduling algorithms are equivalent in semantics.
The basic idea is that positions of load intrinsics and store intrinsics during scheduling should not change the final values of elements.
\tool{} implements this idea by simply separating the memory allocated for load intrinsics and store intrinsics during code generation.
This implementation ensures that memory is not modified by store intrinsics before the memory is accessed by load intrinsics.

\begin{algorithm}[t]
\caption{\rv{All-in Intrinsic Scheduling}}\label{alg:scheduling-allin}
\begin{algorithmic}[1]
\Require Prefix intrinsics $P$, suffix intrinsics $S$, operation intrinsic $I$, common size $N$.
\Function{Scheduling\_AllIn}{$P, S, I, N$}
\State res $\gets$ [ ]
\For{$i \gets 0$ to $N-1$ }
\ForAll{$p \in P[i]$}
\State res.push\_back($p$)
\EndFor
\EndFor
\For{$i \gets 0$ to $N-1$ }
\State res.push\_back($I[i]$)
\EndFor
\For{$i \gets 0$ to $N-1$ }
\ForAll{$s \in S[i]$}
\State res.push\_back($s$)
\EndFor
\EndFor
\State \Return res
\EndFunction

\end{algorithmic}
\end{algorithm}

\begin{algorithm}[t]
\caption{\rv{Unit Intrinsic Scheduling}}\label{alg:scheduling-unit}
\begin{algorithmic}[1]
\Require Prefix intrinsics $P$, suffix intrinsics $S$, operation intrinsic $I$, common size $N$.

\Function{Scheduling\_Unit}{$P, S, I, N$}
\State res $\gets$ [ ]
\For{$i \gets 0$ to $N-1$ }
\ForAll{$p \in P[i]$}
\State res.push\_back($p$)
\EndFor
\State res.push\_back($I[i]$)
\ForAll{$s \in S[i]$}
\State res.push\_back($s$)
\EndFor
\EndFor
\State \Return res
\EndFunction

\end{algorithmic}
\end{algorithm}

\begin{algorithm}[t]
\caption{\rv{Random Intrinsic Scheduling}}\label{alg:scheduling-random}
\begin{algorithmic}[1]
\Require Prefix intrinsics $P$, suffix intrinsics $S$, operation intrinsic $I$, common size $N$.

\Function{Scheduling\_Random}{$P, S, I, N$}
\State res $\gets$ [ ]
\State ptr\_op $\gets $ res.begin() \Comment{point to the last inserted op intrinsic}
\For{$i \gets 0$ to $N-1$ }
\State ptr\_op $\gets $ randomInsert(ptr\_op, res.end(), $I[i]$) 
\State ptr\_begin $\gets $ res.begin(), ptr\_end $\gets $ ptr\_op
\ForAll{$p \in P[i]$}
\State ptr\_begin $\gets $ randomInsert(ptr\_begin, ptr\_end, $p$)
\EndFor
\State ptr\_begin $\gets $ ptr\_op, ptr\_end $\gets $ res.end()
\ForAll{$s \in S[i]$}
\State ptr\_begin $\gets $ randomInsert(ptr\_begin, ptr\_end, $s$)
\EndFor
\EndFor
\State \Return res
\EndFunction

\end{algorithmic}
\end{algorithm}

\subsection{Code Generation}\label{sec:3.4}
After a complete invocation sequence of \rvi{} in the preceding steps, there are four steps left to get a valid C program with \rvi{}.
(1) Global variable declaration. \tool{} uses a series of global variables as the allocated memory to be processed.
Each vector register \texttt{vreg\_mem} (in Algorithm~\ref{alg:register-allocation}, Line 7) corresponds to a global variable as well as a load intrinsic.
(2) Memory initialization. Before operating \rvi{}, all global variables are initialized by randomly generated values of the corresponding type.
To avoid potential floating-point problems such as floating-point precision, \tool{} generates an unsigned integer value in corresponding bits and converts the value to a float by a union data structure. 
How \tool{} initializes scalars in memory is shown in Table~\ref{table:datagen}.
(3) Loop generation. Statements that initialize and update loop variables are added.
(4) Memory print. After all operations are finished, programs generated by \tool{} output the final values of all non-agnostic elements in memory for differential testing.
We discuss how to judge whether an element is non-agnostic in the Section~\ref{sec:3.5}.

\begin{table}[t]
    \centering
    \vspace{-0.2cm}
    \caption{Scalar data generation in \tool{}.}
    \label{table:datagen}
\begin{threeparttable}
\begin{tabular}{lll}
\toprule
\textbf{Type}  & \textbf{Bits}          & \textbf{Generation range (n bits)}        \\
\midrule
bool  & 1             & {[}$0$, $1${]}                  \\
int   & 8, 16, 32, 64 & {[}$-2^{n-1}$, $2^{n-1}-1${]}                   \\
uint  & 8, 16, 32, 64 & {[}$0$, $2^{n}-1${]}                  \\
float & 8, 16, 32, 64 & uint2float\_binary(uint(n))$^{*}$ \\
\bottomrule
\end{tabular}
\begin{tablenotes}
\item[*] If the return value is NaN, generate zero.
\end{tablenotes}
\end{threeparttable}
\end{table}

\subsection{Avoiding Undefined Behaviors}\label{sec:3.5}
General undefined behaviors of C programs are discussed in previous work, such as Csmith~\cite{Csmith}.
In this section, we focus on the undefined behaviors that are related to \rvi{}.
\tool{} avoids all the following undefined behaviors.

\textbf{Masked-off elements and tail elements.} \rvi{} use the masking mechanism to represent whether an element should be executed (i.e., an implementation of control flow).
Masked-off elements are those with a zero mask, indicating that the elements do not need to be executed.
Tail elements refer to the unused elements in vector registers. For example, if a vector register can hold eight elements but only six are used, the remaining two elements are considered tail elements.
For mask-agnostic intrinsics and tail-agnostic intrinsics, values of masked-off elements and tail elements are unknown (i.e., non-agnostic).
\tool{} is implemented with an agnostic-state model to label active elements (i.e., elements neither masked-off nor tail) and generates print statements for only active elements.
In the agnostic-state model, an element is non-agnostic if and only if this element is active and all source elements are non-agnostic.
\tool{} ensures that every printed element is well defined.

\textbf{Conditionally undefined intrinsics.} Some \rvi{} conditionally return uninitialized values for active elements.
\texttt{vrgather} intrinsics, \texttt{vslide} intrinsics, \texttt{vcompress} intrinsics, \texttt{vcpop} intrinsics, \texttt{vfirst} intrinsics, \texttt{vmsif} intrinsics, \texttt{vmsbf} intrinsics, \texttt{vmsof} intrinsics, and \texttt{viota} intrinsics are all conditionally undefined intrinsics as found by us.
For intrinsics of this type, \tool{} uses a rule-based technique to generate ``absolutely correct'' values instead of random values in Table~\ref{table:datagen} as arguments to ensure that the return values are well defined (i.e., no uninitialized values).
The specific rule of data generation depends on the semantic information of each intrinsic.

\textbf{Intrinsics that are always undefined.}
Some \rvi{} always return uninitialized values for active elements.
All values returned by \texttt{vundefined} intrinsics are uninitialized.
The values in the extended portion from \texttt{vlmul\_ext} intrinsics are uninitialized.
It is difficult to determine whether the values returned by \texttt{vreinterpret} intrinsics are well defined.
For intrinsics of this type, \tool{} removes the returned vector register from the \texttt{VregTable} during data flow construction to prevent contamination of subsequent elements.
\tool{} also does not generate print statements for values returned by these intrinsics during code generation.

\textbf{Array safety.} To avoid the problem of array index out-of-bound, programs generated by \tool{} adhere to the programming specifications of \rvi{}.
Arguments of ``\texttt{size vl}'' parameters are returned from \texttt{vsetvl} intrinsics.
Arguments of index parameters in indexed load/store intrinsics are returned from \texttt{vid} intrinsics.
Fine-grained adjustments are applied to the arguments to ensure array safety.
Compilers for \rvi{} do not perform bound checking, which is known as the correctness-security gap~\cite{SP_gap}.
This undefined behavior can lead to security problems such as out-of-bound write.
We provide an out-of-bound write case involving \rvi{}; this case is found and fixed during the development of \tool{} and reported in \#\href{https://github.com/llvm/llvm-project/issues/117677}{117677}.
This out-of-bound write is caused by the absence of adjustments for the \texttt{vl} arguments of segment load/store intrinsics.
After continuous communication with RISC-V officials and improvements to the program generation approach used in \tool{}, no out-of-bound arrays currently exist in the programs generated by \tool{}.

\textbf{Numerical safety.}
The data generation approach of \tool{} ensures that the initial values of elements are valid and safe.
Integer overflow and NaN (Not-a-Number) values may occur during computation by \rvi{}.
\tool{} converts NaN to a specific valid value only during data generation and print-statement generation.
The behavior of integer overflow in RVV intrinsics is well defined by the documentation, and no additional operations are required.
Elements are used for only computation to avoid causing follow-up issues (e.g., infinite loop).
\tool{} does not ensure that every element is valid during computation but ensures that invalid elements do not cause issues or inconsistencies in printed values.

\section{Evaluation}\label{sec:evaluation}
This section presents the details of our evaluation of \tool{}.
The effectiveness of \tool{} is mainly evaluated from three dimensions: bug detection, coverage (including code coverage and intrinsic coverage), and performance.

\subsection{Experimental Setup}\label{sec:4.1}

\textbf{Compilers under test.}
Mainstream compilers for \rvi{} are shown in Table~\ref{table:compilers}. 
Our experiments are limited to compilers that support the ratified \rvi{}. 
GCC ($\geq 14.1.0$)~\cite{GCC} and LLVM ($\geq 17.0.1$)~\cite{LLVM}, the most popular open-source compilers, both support ratified \rvi{}.
XuanTie~\cite{XuanTie} is a compiler developed by Alibaba DAMO Academy based on GCC and LLVM to match XuanTie CPUs (e.g., C906), and also support ratified \rvi{}.
Compilers that implement only draft versions of \rvi{} shown in Table~\ref{table:compilers} or lack the support of \rvi{} (including old versions of GCC, LLVM, and XuanTie) are excluded from our experiments.
XuanTie is excluded from code coverage evaluation, as its source code is not available.

\begin{table}[t]
    \centering
    \vspace{-0.2cm}
    \caption{Mainstream compilers for \rvi{}.}
    \label{table:compilers}
\begin{threeparttable}
\begin{tabular}{llll}
\toprule
RVV Spec            & GCC           & LLVM        & XuanTie       \\ \midrule
Draft ($\leq$ 0.11)       & 13            & 16          & gcc-v2             \\
Ratified (0.12\&1.0)    & 14, 15$^{*}$  & 17, 18, 19, 20$^{*}$ & gcc-v3, llvm-v2    \\
\bottomrule
\end{tabular}
\begin{tablenotes}
\item[*] Experimental version.
\end{tablenotes}
\end{threeparttable}
\end{table}

\rv{\textbf{Compiler flags.} 
We use all five standard optimization flags, i.e., \texttt{-O0}, \texttt{-O1}, \texttt{-O2}, \texttt{-O3}, and \texttt{-Os}, for fuzzing compilers in the latest released versions and experimental versions.
In the experiments involving historical versions and performance analysis, only the \texttt{-O0} and \texttt{-O3} optimization flags are used.
Other related compiler flags are set as follows: "\texttt{-march=rv64gcv\_zvfh -mabi=lp64d -Wno-psabi -static}".}

\textbf{Environment.}
We conduct all our evaluations on a docker container running Ubuntu 24.04.1 LTS in a Linux server.
The Linux server is equipped with two AMD EPYC 7H12 64-Core CPUs and each CPU has 512GB RAM.
RISC-V ELF files after compilation are executed by QEMU in version 9.1.0.


\subsection{Quantitative Bug-Finding Results} \label{sec:4_quantitative}
This subsection presents various summary statistics on the results of our compiler bug detection effort.

\begin{table}[t]
    \centering
    \vspace{-0.2cm}
    \caption{Type and status of new compiler bugs.}
    \label{table:bug_type}
\begin{tabular}{lccc|c}
\toprule
\textbf{Symptom}        & \textbf{GCC} & \textbf{LLVM} & \textbf{XuanTie} & \textbf{Total} \\ \midrule
Compiler Crash & 2   & 0    & 1           & 3     \\
Runtime Crash & 2   & 1    & 2           & 5     \\
Wrong Result  & 3   & 0    & 2           & 5     \\ \midrule
\textbf{Total} & 7   & 1    & 5           & 13    \\
(Confirmed | Fixed)      & (6|1)   & (0|1)    & (4|1)           & (10|3)   \\
\bottomrule
\end{tabular}
\end{table}

\textbf{Number of bugs.} Table~\ref{table:bug_type} summarizes 13 previously unknown bugs uncovered by \tool{} in the latest released versions and experimental versions of compilers under test.
All the bugs are confirmed as real-world bugs by the corresponding compiler developers, and three bugs are fixed.
Six unrepaired bugs of GCC are labeled target milestones, which define when the bugs are expected to be fixed.
\tool{} detects only one bug in the latest version (19.1.4 by the time that we experiment) of LLVM, and this bug is detected independently by the LLVM developers nearly simultaneously.
Five bugs are detected in the latest version of XuanTie, and we report these bugs by emailing XuanTie developers.
Three bugs lead to compiler crashes when parsing \rvi{} (two null-pointer exceptions and one floating-point exception).
Five bugs lead to runtime crashes by generating illegal instructions or trying to access unavailable memory.
Five bugs lead to incorrect calculation results.

\rv{
\textbf{Comparison with baselines.}
To demonstrate \tool{}'s superiority in bug detection, we compare its results with two baselines, RIF and Csmith, by measuring the number of detected bugs.
During a one-week experiment to fuzz the latest versions of GCC (14.2.0), LLVM (19.1.4), and XuanTie (gcc-v3, llvm-v2), the baseline tools (RIF and Csmith) do not detect any bug, while \tool{} identifies multiple bug cases. 
As the only tool that supports \rvi{} before \tool{}, RIF cannot detect bugs for two reasons.
First, RIF is implemented based on the draft specification rather than the ratified specification, leading to numerous compilation errors when compiling generated programs as the RVV specification is updated.
Second, as discussed in Section~\ref{sec:intro}, RIF supports extremely limited \rvi{} and does not support operation combinations in a loop, resulting in an inherent limitation in bug detection. 
}

\rv{
\textbf{Security impacts.}
Using the programs generated by \tool{} and our differential testing framework, \tool{} can detect not only compiler crashes but also security bugs.
The impacts of these security bugs include unintended rounding (e.g., \#\href{https://gcc.gnu.org/bugzilla/show_bug.cgi?id=118103}{118103}), data loss (e.g., \#\href{https://gcc.gnu.org/bugzilla/show_bug.cgi?id=117947}{117947}), illegal memory access (e.g., \#\href{https://gcc.gnu.org/bugzilla/show_bug.cgi?id=118100}{118100}), etc.
These bugs are difficult to detect because they occur under seemingly normal conditions, without any compilation warning/error messages.
These bugs can lead to serious security issues when bug-triggering SIMD intrinsics are used to accelerate the processing of sensitive data, such as quantitative computations in financial systems.
}

\textbf{Affected \rvi{}.} Some bugs detected by \tool{} are related to specific intrinsics. By manually debugging bug cases, we identify the affected \rvi{} of these bugs.
Other bugs are caused by overly complex combinations of \rvi{}, making it difficult for us to identify the specific affected \rvi{}.
Table~\ref{table:bug_affect} summarizes the known affected \rvi{}.
Most of non-compiler-crash bugs are caused by faults related to CSRs, especially CSRs that are explicitly controlled by corresponding parameters in intrinsics (e.g., vxrm and frm).
Although the number of unique bugs is limited, a significant number of \rvi{} are affected by these bugs and have the potential to trigger them.
More than 20,000 \rvi{} are known to be affected by new bugs detected by \tool{}.
The significant number of affected \rvi{} reflects the importance of the detected bugs.

\begin{table}[t]
    \centering
    \footnotesize
    \vspace{-0.2cm}
    \caption{Number of known affected \rvi{}.}
    \label{table:bug_affect}

\begin{tabular}{llll}
\toprule
\textbf{Category}                                              & \textbf{Symptom}        & \textbf{Compiler}         & \textbf{\#N}              \\ \midrule
LMUL Extension Intrinsics                      & Compiler Crash & GCC, XuanTie & 270                   \\  \midrule
LMUL Truncation Intrinsics                     & Compiler Crash & GCC, XuanTie & 270                   \\  \midrule
Vector Insertion Intrinsics                           & Compiler Crash & GCC              & 292                   \\  \midrule
Fixed-Point Intrinsics with `vxrm'         & Wrong Result  & GCC              & 4416                  \\  \midrule
Masked Widening Intrinsics                            & Runtime Crash & XuanTie      & 7376     \\  \midrule
\multirow{2}{*}{Floating-Point Intrinsics with `frm'} & Runtime Crash & LLVM             & \multirow{2}{*}{9020} \\ 
                                                      & Wrong Result  & GCC, XuanTie &                       \\
\bottomrule
\end{tabular}
\end{table}

\begin{figure*}[t]
    \centering
    \begin{minipage}{\textwidth}
        \begin{minipage}[b]{0.405\textwidth}
            \centering
            \includegraphics[width=0.98\textwidth]{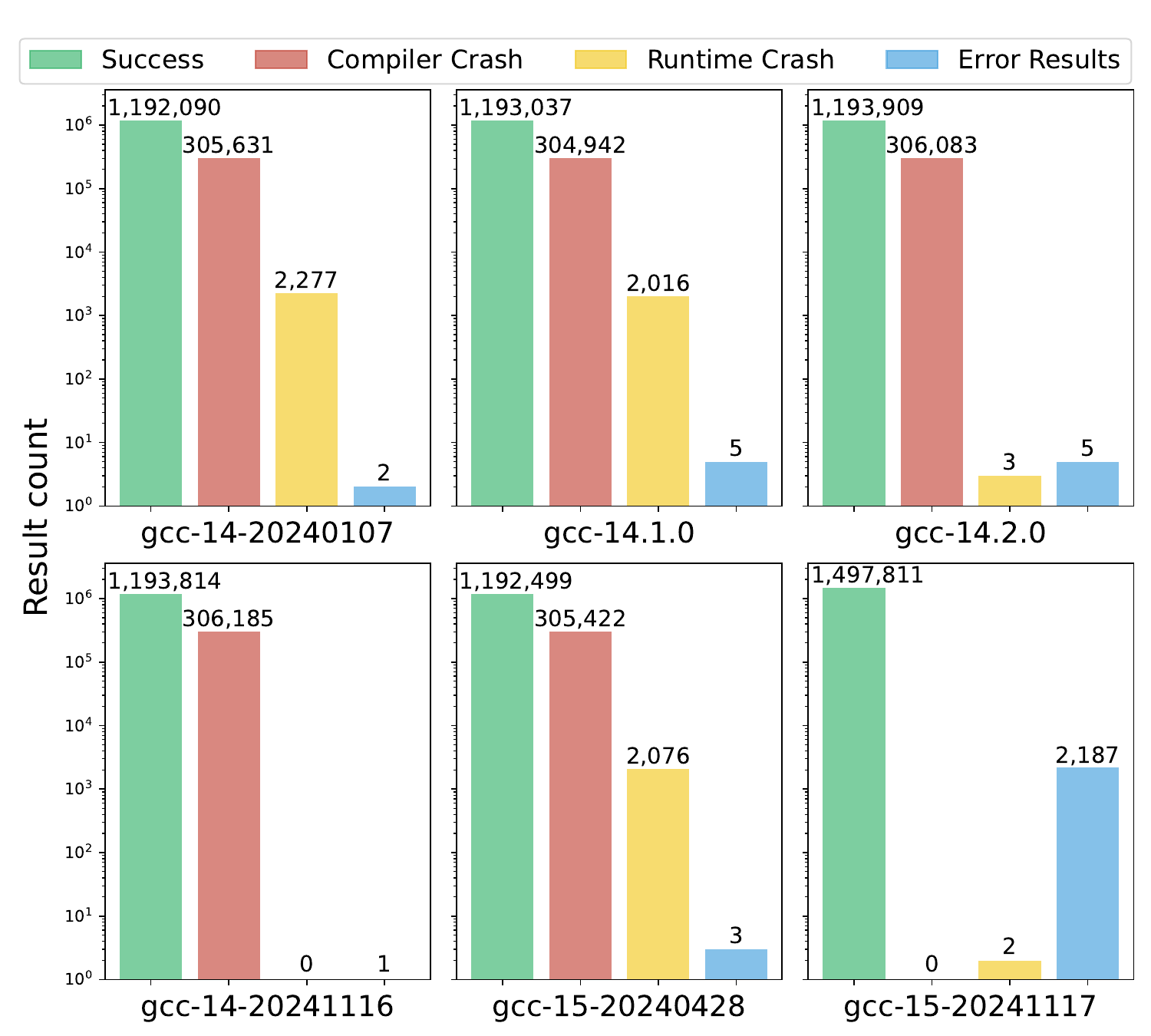}
            \vspace{-0.25cm}
            \caption*{(a) GCC}
    \end{minipage}
    \begin{minipage}[b]{0.568\textwidth}
            \centering
            \includegraphics[width=0.98\textwidth]{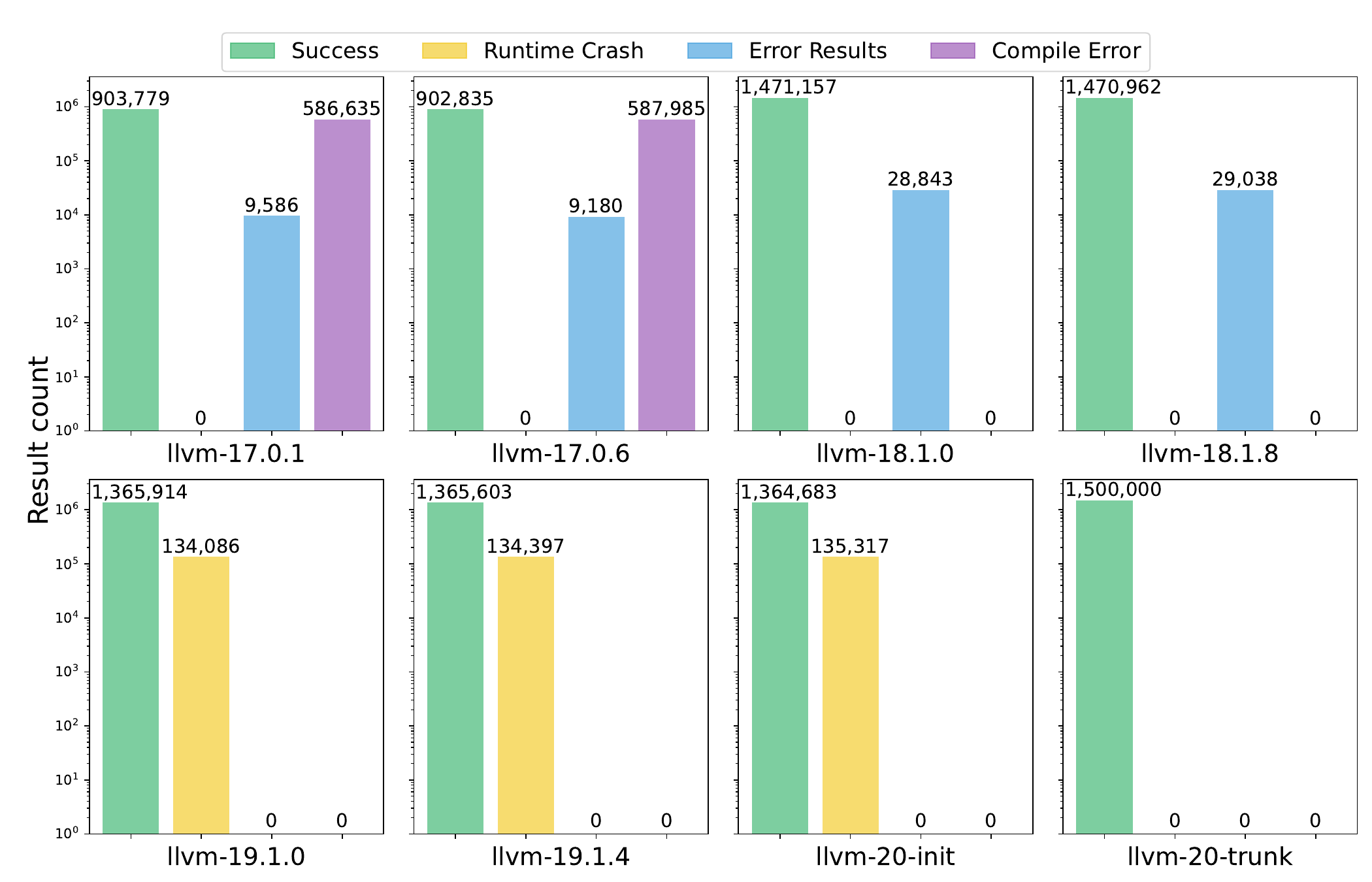}
            \vspace{-0.25cm}
            \caption*{(b) LLVM}
        \end{minipage}
    \end{minipage}
    \vspace{-0.25cm}
    \caption{Quantitative comparison results of GCC and LLVM versions.}
    \label{fig:gcc&llvm}
\Description[<short description>]{<long description>}
\end{figure*}

\textbf{Quantitative comparison of GCC and LLVM versions.} Figure~\ref{fig:gcc&llvm} shows the results of our quantitative comparison experiments.
For each major version of GCC and LLVM shown in Table~\ref{table:compilers}, we include the earliest and latest available versions.
As GCC has limited versions that support ratified \rvi{}, we also include all released versions of GCC (14.[1-2].0).
For each compiler under test, we use \tool{} to randomly generate 1,500,000 programs (500,000 seeds * 3 scheduling algorithms), and each program is compiled at \texttt{-O0} and \texttt{-O3}.
The data length is randomly selected from [1, 1000], and the sequence length is randomly selected from [1, 100].

By the quantitative comparison of GCC and LLVM versions, we derive the following findings.
First, compiler bugs related to \rvi{} exist widely in GCC and LLVM.
Among the tested versions, there is no detected bugs for LLVM-20-trunk, as the only detected LLVM bug \#\href{https://github.com/llvm/llvm-project/issues/117909}{117909} illustrated in Figure~\ref{fig:bug_illegal_instruction} has been fixed.
Second, \tool{} is also capable of detecting many fixed bugs that are not included in Table~\ref{table:bug_type}. 
For example, a large number of programs compiled by LLVM-17 and LLVM-18 lead to different results between \texttt{-O0} and \texttt{-O3}; however, this issue does not occur with LLVM-19 and LLVM-20.
Third, programs that are capable of triggering the same bug are diverse.
This finding indicates that a substantial number of intrinsics are affected by the bugs, being consistent with our previous discussion.

We provide explanations for some details in Figure~\ref{fig:gcc&llvm}.
Figure~\ref{fig:gcc&llvm}(b) does not include the ``Compiler Crash'' column because \tool{} does not detect any compiler crash in LLVM.
A large number of programs fail to compile with LLVM-17 because it does not support certain ratified \rvi{}.

\rv{\textbf{Findings on differential-testing strategies.}
As discussed at the beginning of Section~\ref{sec:approach}, three differential-testing strategies are used, comparing the results from (1) different compilers, (2) different optimizations, and (3) equivalent programs.
In our experiments, all previously unknown bugs can be detected by the cross-compiler and cross-optimization strategies.
However, we find that comparing results from equivalent programs compiled by a single compiler with a single optimization detects only four previously unknown bugs, and all these bugs are compiler crashes and runtime crashes.
This finding provides us with three key insights.
First, differential testing between compilers and optimizations is crucial.
Most security bugs and logical bugs, rather than crashes, are related to the optimization components in compilers.
Second, increasing diversity in invocation sequences by introducing various intrinsic-scheduling algorithms is effective.
We observe that some compiler/runtime crashes are triggered by only specific scheduling algorithms.
For example, the bug in \#\href{https://github.com/llvm/llvm-project/issues/117909}{117909} cannot be triggered by the unit scheduling algorithm.
Third, the current construction of equivalent programs is not critical.
We do not detect more new bugs by the current equivalence, and more effective approaches for the construction of equivalent programs may help to address this limitation.}
\subsection{Coverage}\label{sec:4_coverage}

\textbf{Code coverage.} We conduct code coverage analysis to evaluate the effectiveness of \tool{} in improving code coverage.
We use \tool{} to generate 10,000 random programs (10,000 random seeds * 1 scheduling mode) including all four categories of \rvi{}.
For comparison against \tool{}, we also use Csmith (a well-known code generator for compiler testing) and RIF (the only compiler fuzzer for \rvi{}) as baselines.
We use Csmith and RIF to generate 10,000 random programs respectively.
For the sake of fairness, the data lengths and operation lengths of RIF and \tool{} are set to 10.
We use the -O3 optimization flag when compiling programs generated by Csmith, RIF, and \tool{}.
We use \texttt{gcov} for collecting GCC code coverage and \texttt{llvm-cov} for collecting LLVM code coverage.
We control the number of test cases rather than the duration of fuzzing, as the generation time proportion of code generators is small and the most time-consuming parts are compilation and execution, which are proportional to the number of test cases (we discuss the performance details in Section~\ref{sec:4_performance}).
Based on previous experiments in compiler testing~\cite{Csmith, pldi24-inject}, compiling 10,000 random programs that are generated by each generator is sufficient to achieve meaningful code coverage results.

Code coverage results of GCC and LLVM source code are shown in Table~\ref{table:code-coverage}, and we have two main findings.
First, \tool{} achieves significantly higher code coverage than using only existing code generators Csmith and RIF.
For example, compared to compiling 10,000 programs generated by Csmith and 10,000 programs generated by RIF, compiling other 10,000 programs generated by \tool{} can remarkably increase line coverage in GCC by 11.81\% (124,948 more lines) and in LLVM by 5.78\% (91,987 more lines).
Second, employing these generators in a complementary manner achieves more code coverage than relying on a single generator alone.
This finding indicates that each generator uniquely covers some specific portions of the GCC and LLVM source code.

\begin{table}[t]
    \centering
    \small
    \vspace{-0.2cm}
    \caption{Function coverage (FC), line coverage (LC), and branch coverage (BC) of GCC and LLVM source code.}
    \label{table:code-coverage}
\begin{tabular}{lllll}
\toprule
\multicolumn{1}{l}{} & \textbf{Generator}               & \textbf{FC} & \textbf{LC} & \textbf{BC} \\ \midrule
\multirow{10}{*}{GCC}        & RVISmith                &26.88\%    &23.77\%  &14.90\%    \\ \cline{2-5}      
                             & Csmith                  &14.42\%    &12.08\%  &6.25\%  \\
                             & Csmith+RVISmith         &27.13\%    &24.37\%  &15.21\%    \\ 
                             & (absolute change) &+16,011 &+130,202 &+147,470 \\ \cline{2-5}
                             & RIF                     &6.05\%    &6.18\%    &3.69\%    \\
                             & RIF+RVISmith            &27.51\%    &24.39\%    &15.20\%    \\ 
                             & (absolute change) &+27,061 &+192,924 &+189,344 \\ \cline{2-5}
                             & Csmith+RIF              &15.74\%    &12.95\%    &6.64\%    \\
                             & Csmith+RIF+RVISmith     &27.63\%    &24.76\%    &15.39\%    \\ 
                             & (absolute change) &+14,985 &+124,948 &+143,891 \\ \midrule
\multirow{10}{*}{LLVM}       & RVISmith                &21.51\%     &14.46\%     &13.64\%    \\ \cline{2-5}
                             & Csmith                  &16.00\%    &8.82\%    &6.79\%    \\
                             & Csmith+RVISmith         &21.98\%    &14.91\%     &14.10\%    \\
                             & (absolute change) & +7,686 & +96,872 & +70,114 \\ \cline{2-5}
                             & RIF                     &8.76\%     &4.75\%    &5.05\%    \\
                             & RIF+RVISmith            &22.29\%    &15.05\%    &14.26\%    \\ 
                             & (absolute change) & +17,398 & +164,001 & +88,336 \\ \cline{2-5}
                             & Csmith+RIF              &16.90\%    &9.50\%    &8.20\%    \\
                             & Csmith+RIF+RVISmith     &22.49\%    &15.28\%    &14.48\%    \\ 
                             & (absolute change) & +7,196 & +91,987 & +60,221 \\
\bottomrule
\end{tabular}
\end{table}

\textbf{Intrinsic coverage.}
\tool{} focuses on compilers for \rvi{}. 
However, code coverage of the entire source code may not reflect the extent of testing relevant compiler portions that support \rvi{}.
GCC and LLVM are huge code projects that contain a substantial amount of complex code unrelated to \rvi{}, as the two compilers support multiple source languages, backends, and configuration options.

To address the aforementioned limitation of code coverage, we introduce the intrinsic coverage metric to measure the effectiveness of a code generator in covering \rvi{}.
Intrinsic coverage provides insights into how thoroughly \rvi{} are being tested, helping identify untested parts of \rvi{}.
How we calculate intrinsic coverage is shown in the following formula.
In the formula, $count_i$ represents the count of the appearances of $i$-th intrinsic's name in the generated code, and $weight_i$ represents the count of the $i$-th intrinsic's name in the list of intrinsic definitions.
Note that $weight_i=1$ for non-overloaded intrinsics, and $weight_i>1$ for overloaded intrinsics.
By this formula, we can obtain approximate intrinsic coverage with simple static analysis of intrinsics names instead of developing a complex tool to find out which overloaded intrinsic is invoked in a program.
\rv{
This approximation is intentionally designed as overloaded intrinsics (the same SEW/LMUL ratio and the same operation) have an equal probability of coverage by \tool{}.
}

\begin{align*}
Intrinsic \ coverage = \frac{ \sum_{i=0}^{n} \min(count_i, weight_i) }{ \sum_{i=0}^{n} weight_i }
\end{align*}

\begin{table*}[t]
    \centering
    \vspace{-0.2cm}
    \caption{Intrinsic coverage by RIF and RVISmith.}
    \label{table:intrinsic-coverage}
\begin{tabular}{lll|ll|ll|ll}
\toprule
\multirow{2}{*}{\begin{tabular}[c]{@{}l@{}}Intrinsic \\ under test\end{tabular}}        & \multicolumn{2}{c}{$n=10^{2}$} & \multicolumn{2}{c}{$n=10^{3}$} & \multicolumn{2}{c}{$n=10^{4}$} & \multicolumn{2}{c}{$n=10^{5}$} \\ \cline{2-9} 
                 & RIF      & RVISmith              & RIF      & RVISmith               & RIF      & RVISmith       & RIF      & RVISmith          \\ \midrule
Explicit         & 5.11\%   & 6.90\% (+1.79\%)      & 14.21\%  & 32.09\% (+17.88\%)     & 15.18\%  & 65.25\% (+50.07\%)     & 15.18\%  &  78.69\% (+63.51\%) \\
Explicit, policy & 0.75\%   & 4.61\% (+3.86\%)      & 5.62\%   & 28.22\% (+22.60\%)     & 10.59\%  & 69.92\% (+59.33\%)     & 10.60\%  &  70.79\% (+60.19\%) \\
Implicit         & 0.00\%   & 7.27\% (+7.27\%)      & 0.00\%   & 42.14\% (+42.14\%)     & 0.00\%   & 66.50\% (+66.50\%)     & 0.00\%   &  78.65\% (+78.65\%) \\
Implicit, policy & 0.00\%   & 4.85\% (+4.85\%)      & 0.00\%   & 34.77\% (+34.77\%)     & 0.00\%   & 70.44\% (+70.44\%)     & 0.00\%   &  70.44\% (+70.44\%) \\
Total            & 1.35\%   & 5.74\% (+4.39\%)      & 4.76\%   & 33.84\% (+29.08\%)     & 6.39\%   & 68.32\% (+61.93\%)     & 6.39\%   &  74.08\% (+67.69\%) \\    
\bottomrule
\end{tabular}
\end{table*}

\begin{figure}
  \begin{center}
    \includegraphics[width=0.47\textwidth]{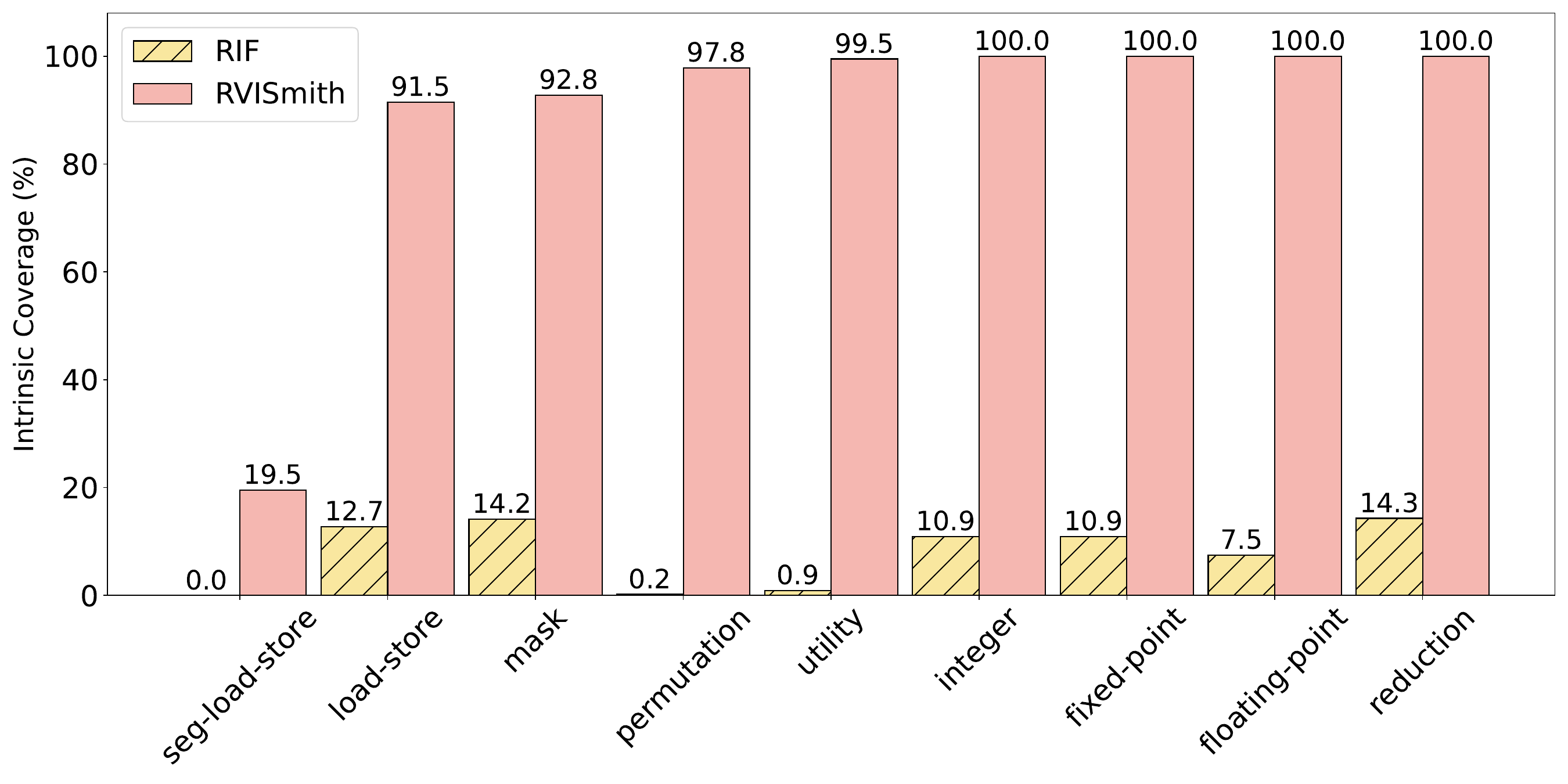}
  \end{center}
  \vspace{-0.2cm}
  \caption{Distribution of covered RVV intrinsics by RIF and \tool{} ($n=10^{5}$).} 
  \label{fig:cov-dis}
\Description[<short description>]{<long description>}
\end{figure}

We use RIF and \tool{} to generate random programs with $n$ random seeds, where the data lengths and operation lengths are set to 10.
Table~\ref{table:intrinsic-coverage} summarizes the experimental results of RIF and \tool{} in the intrinsic coverage metric.
Csmith is excluded from this experiment because it cannot generate any \rvi{}.
The results show that \tool{} achieves significantly higher intrinsic coverage than RIF across all four categories of \rvi{} being tested.
\tool{} supports all four categories of \rvi{}; however, RIF supports only explicit intrinsics (with and without policy).
\rv{
There is little difference in coverage by \tool{} between explicit (non-overloaded) and implicit (overloaded) intrinsics when test cases are enough ($n>=10^{4}$).
}
Compared to RIF, \tool{} totally achieves 11.5 times higher intrinsic coverage (from 6.39\% to 74.08\% when $n$ is $10^{5}$).

Figure~\ref{fig:cov-dis} shows the distribution of \rvi{} covered by RIF and \tool{}, with \rvi{} classified by functionality according to the RVV intrinsic document.
Compared to \tool{}, the intrinsic coverage of RIF is low for any category of \rvi{}.
\tool{} achieves over 90\% intrinsic coverage on all \rvi{} except segment load/store intrinsics.
We provide the reasons why \tool{} achieves low coverage on segment load/store intrinsics in the experiments.
First, the number of segment load/store intrinsics is large. 
The number of segment load/store intrinsics exceeds 37,000, making it the most numerous among all types.
Second, the \tool{} approach has a low probability of generating programs including segment load/store intrinsics, as the number of operation intrinsics capable of reading/writing data via segment load/store intrinsics is much smaller than the total number of segment load/store intrinsics. 
This low probability results in limited segment load/store intrinsics being inserted during intrinsic scheduling.
A coverage-guided fuzzing approach for RVV intrinsic compilers can be explored in future work to address this limitation of \tool{}.


\subsection{Performance of \tool{}}\label{sec:4_performance}
We measure the distribution of CPU time and real time used by each main step in our pipeline when fuzzing on the latest released versions of GCC (14.2.0) and LLVM (19.1.4).
CPU time is collected by \texttt{time.process\_time()}, and real time is collected by \texttt{time.time()}.
\tool{} is used in its default configuration for explicit intrinsics.
Our hardware configuration is discussed in Section~\ref{sec:4.1}.
We use \texttt{ProcessPoolExecutor} for parallel processing, and all CPU cores are used.
As shown in the results in Table~\ref{table:performance}, the time for \tool{} to generate test cases accounts for a small proportion of both CPU time (0.021\%) and real time (0.022\%).
Most of the time is spent on compilation and execution.
This finding indicates that the performance of \tool{} is not a bottleneck when fuzzing GCC and LLVM.
The performance analysis also reveals other interesting findings.
For example, the proportion of CPU time for GCC and LLVM is similar, but GCC takes approximately twice as much real time as LLVM when compiling test cases.

\begin{table}[t]
    \centering
    \vspace{-0.2cm}
    \caption{CPU time and real time proportion when fuzzing on the latest released versions of GCC and LLVM ($n=10^{5}$).}
    \label{table:performance}
\begin{tabular}{llrr}
\toprule
\textbf{Tool}      & \textbf{Step}     & \textbf{CPU Time}  & \textbf{Real Time} \\ \midrule
\tool{}   & generation  &  0.021\%    &0.022\%                 \\ \midrule
gcc \texttt{-O0}   & compilation &  13.120\%   &33.700\%                 \\
qemu      & execution   &  13.317\%   &0.228\%                 \\ \midrule
gcc \texttt{-O3}   & compilation &  13.715\%   &33.219\%                 \\
qemu      & execution   &  12.994\%   &0.213\%                 \\ \midrule
clang \texttt{-O0} & compilation &  11.665\%   &15.295\%                 \\
qemu      & execution   &  11.619\%   &0.233\%                 \\ \midrule
clang \texttt{-O3} & compilation &  11.945\%   &16.858\%                 \\
qemu      & execution   &  11.605\%   &0.232\%                 \\ 
\bottomrule
\end{tabular}
\end{table}

\subsection{Case Study}\label{sec:4_bugcase}

\tool{} is capable of detecting compiler bugs related to \rvi{}.
We discuss a selection of bugs discovered by \tool{}.

\textbf{Data loss.} Figure~\ref{fig:bug_example_dataloss} shows a program that triggers a miscompilation bug of GCC.
In each iteration, two vectors with \texttt{vl} elements each are added by the \texttt{vadd\_vv} intrinsic.
However, when the program is compiled by GCC at -O2/-O3 optimizations, pointers to the allocated memory are updated by the CSR \texttt{vlenb} that is incorrectly configured by a \texttt{vsetvli} instruction.
This miscompilation causes the \texttt{vl} elements to be successfully added, while the next \texttt{vl} elements are skipped in each iteration.
Finally, half of the data are lost in the computation result.

\begin{figure}[t]
    \centering
\begin{minipage}[b]{0.45\textwidth}
\begin{minted}[fontsize=\small, highlightlines={2,11}, highlightcolor=yellow]{c}
for (size_t vl; avl > 0; avl -= vl){
  vl = __riscv_vsetvl_e64m8(avl);
  vint8m1_t mask_value = __riscv_vle8_v_i8m1(ptr_mask, vl);
  vbool8_t vmask=__riscv_vmseq_vx_i8m1_b8(mask_value,1,vl);
  vuint8m1_t va = __riscv_vle8_v_u8m1(ptr_a, vl);
  vuint8m1_t vb = __riscv_vluxei8_v_u8m1(ptr_b, \
  __riscv_vsll_vx_u8m1(__riscv_vid_v_u8m1(vl), 0, vl), vl);
  vuint8m1_t vc = __riscv_vadd_vv_u8m1_m(vmask,va,vb,vl);
  __riscv_vse8_v_u8m1(ptr_c, vc, vl);
  /*some other intrinsics (ratio=8)*/
  ptr_mask += vl; ptr_a += vl; ptr_b += vl; ptr_c += vl;
}
\end{minted}
\end{minipage}
    \vspace{-0.2cm}
    \caption{[\#\href{https://gcc.gnu.org/bugzilla/show_bug.cgi?id=117947}{117947}] GCC at \texttt{-O2/3} miscompiles this code. Pointers are updated by the CSR \texttt{vlenb} that is incorrectly configured by a \texttt{vsetvli} instruction. This bug results in half of the data being lost after computation.}
    \label{fig:bug_example_dataloss}
\Description[<short description>]{<long description>}
\end{figure}

\textbf{Unintended rounding.} Figure~\ref{fig:bug_example_rounding} shows a program that triggers an unintended rounding bug of GCC.
The first floating-point intrinsic (\texttt{vfnmadd\_vv}) is in the RDN (Round Down) mode, and the second floating-point intrinsic (\texttt{vfmsac\_vv}) is in the default RNE (Round to Nearest) mode.
In the correctly compiled assembly code, the status of \texttt{frm} should be changed before \texttt{vfnmadd\_vv} and be restored between \texttt{vfnmadd\_vv} and \texttt{vfmsac\_vv}.
However, GCC misses the instruction that restores the status of \texttt{frm}, leading to incorrect calculation results.
In certain instances of this bug, the disparity between correct and incorrect results is substantial, as a small number of bits in a floating-point number can significantly affect the numerical value.

\begin{figure}[t]
    \centering
\begin{minipage}[b]{0.45\textwidth}
\begin{minted}[fontsize=\small, highlightlines={4,5,6}, highlightcolor=yellow]{c}
for (size_t vl; avl > 0; avl -= vl){
  vl = __riscv_vsetvl_e16m1(avl);
  vfloat16m1_t va = __riscv_vle16_v_f16m1(ptr_a, vl);
  va = __riscv_vfnmadd_vv_f16m1_rm(va, va, va, \
       __RISCV_FRM_RDN, vl);
  va = __riscv_vfmsac_vv_f16m1(va, va, va, vl);
  __riscv_vse16_v_f16m1(ptr_b, va, vl);
  ptr_a += vl; ptr_b += vl;
}
\end{minted}
\end{minipage}
    \vspace{-0.2cm}
    \caption{[\#\href{https://gcc.gnu.org/bugzilla/show_bug.cgi?id=118103}{118103}] GCC at \texttt{-O3} miscompiles this code. The status of CSR \texttt{frm} is not restored between the first and second floating-point intrinsics, leading to an unintended rounding.}
    \label{fig:bug_example_rounding}
\Description[<short description>]{<long description>}
\end{figure}

\textbf{Compiler crash.} A compiler crash (internal compiler error) is always viewed as a compiler bug because compilers should either work successfully or provide warning/error messages for incorrect input programs.
GCC crashes on the program shown in Figure~\ref{fig:bug_compiler_crash}.
The reason for this crash is that GCC tries to use a NULL pointer when parsing the \texttt{vlmul\_ext} intrinsic.
This bug has been fixed after we report it.

\begin{figure}[t]
    \centering
\begin{minipage}[b]{0.45\textwidth}
\begin{minted}[fontsize=\small, highlightlines={4}, highlightcolor=yellow]{c}
for (size_t vl; avl > 0; avl -= vl){
  vl = __riscv_vsetvl_e16m1(avl);
  vfloat16mf2_t va = __riscv_vle16_v_f16mf2(ptr_a, vl);
  vfloat16m1_t vb = __riscv_vlmul_ext_v_f16mf2_f16m1(va);
  ptr_a += vl;
}
\end{minted}
\end{minipage}
    \vspace{-0.2cm}
    \caption{[\#\href{https://gcc.gnu.org/bugzilla/show_bug.cgi?id=117286}{117286}] GCC at \texttt{-O1/2/3/s} crashes on this code. The LMUL extension intrinsic causes a segment fault in GCC.}
    \label{fig:bug_compiler_crash}
\Description[<short description>]{<long description>}
\end{figure}

\textbf{Illegal instruction.} Figure~\ref{fig:bug_illegal_instruction} shows a program that triggers a miscompilation bug of LLVM.
As previously discussed in this section, instructions that set the appropriate status of the \texttt{frm} CSR should be generated when compiling floating-point intrinsics.
This LLVM bug generates a \texttt{fsrmi} instruction that tries to write an illegal value to the \texttt{frm} CSR, leading to a runtime crash.
\tool{} and LLVM developers detect this bug nearly simultaneously, and this bug has been fixed.

We discuss the reason why these bugs cannot be detected by RIF.
The bugs in Figure~\ref{fig:bug_example_dataloss}, Figure~\ref{fig:bug_example_rounding}, and Figure~\ref{fig:bug_illegal_instruction} cannot be detected by RIF, as these bugs are triggered by complex combinations with multiple operation intrinsics and RIF does not support any combination of operation intrinsics.
The bug in Figure~\ref{fig:bug_compiler_crash} cannot be detected by RIF, because LMUL extension intrinsics and other related intrinsics that can trigger this bug are not supported by RIF.

\begin{figure}[t]
    \centering
\begin{minipage}[b]{0.45\textwidth}
\begin{minted}[fontsize=\small, highlightlines={4,5}, highlightcolor=yellow]{c}
for (size_t vl; avl > 0; avl -= vl){
  vl = __riscv_vsetvl_e32m1(avl);
  vfloat32m1_t va = __riscv_vlse32_v_f32m1(ptr_a, 4, vl);
  va = __riscv_vfsqrt_v_f32m1_rm( va, __RISCV_FRM_RNE, vl);
  va = __riscv_vfredosum_vs_f32m1_f32m1(va, va, vl);
  __riscv_vse32_v_f32m1(ptr_b, va, vl);
  ptr_a += vl; ptr_b += vl;
}
\end{minted}
\end{minipage}
    \vspace{-0.2cm}
    \caption{[\#\href{https://github.com/llvm/llvm-project/issues/117909}{117909}] Clang at \texttt{-O0/1/2/3/s} generates an illegal \texttt{fsrmi} instruction on this code.}
    \label{fig:bug_illegal_instruction}
\Description[<short description>]{<long description>}
\end{figure}


\section{Discussion}\label{sec:discussion}
In this section, we present some limitations of \tool{} and multiple possible future directions.

\textbf{Limitations.} The main limitations of \tool{} are divided into three aspects.
First, the generation of the invocation sequence of \rvi{} is purely random.
A coverage-guided approach can be developed in the future to enhance the coverage of specific intrinsics (e.g., segment load/store intrinsics) and combinations of intrinsics.
Second, \tool{} focuses exclusively on combinations of \rvi{} within a single strip-mining loop, while \rvi{} in complex control flows remain untested.
Previous work~\cite{YARPGen-pldi23} has shown multiple compiler bugs related to loop optimizations.
Third, \tool{} applies rule-based data generation for conditionally undefined intrinsics and indexed load/store intrinsics.
To avoid undefined behaviors, \tool{} generates absolutely correct data for partial intrinsics instead of randomized data generation.
This data generation results in certain scenarios remaining not covered.

\rv{
From the perspective of security vulnerability detection, our work still overlooks three types of security vulnerabilities.
First, vulnerabilities in emulators and hardware related to SIMD instructions are overlooked.
While \tool{} is designed for fuzzing compilers for intrinsics and may help uncover such vulnerabilities, specialized fuzzing tools are still necessary.
Second, vulnerabilities that are related to undefined behaviors are overlooked.
\tool{} is intentionally designed to generate well-defined programs to avoid false positives, and no security-related code (to avoid undefined behaviors) is in the generated programs.
However, some compiler-introduced security vulnerabilities arise from incorrect optimizations applied to security-related code~\cite{compiler_bug_study}.
Third, vulnerabilities in scenarios not covered by current \tool{} are overlooked, as discussed in the preceding limitations.
}

\textbf{Future work.}
We propose three possible future directions derived from \tool{}.
(1) Improvements of the approach for generating code with \rvi{}.
Additional approaches could be developed in the future to address the aforementioned limitations of \tool{} to achieve higher coverage and detect more previously unknown bugs than \tool{}.
(2) Fuzzing compilers for SIMD intrinsics in other ISAs.
\tool{} is the first work in academia focused on fuzzing compilers for SIMD intrinsics to the best of our knowledge, and currently only \rvi{} are supported.
Fuzzing compilers for other SIMD intrinsics, such as SSE/AVX for x86~\cite{Intel-Intrinsics-Guide} and Neon for ARM~\cite{Neon}, still requires significant effort.
Although the idea of \tool{} is general, the implementation of fuzzing tools on other ISAs requires solving domain-specific problems.
(3) Fuzzing emulators and hardware for SIMD instructions.
In the experiments of this paper, we use only the latest QEMU v9.1.0 to execute RISC-V ELFs after compilation, and we assume that there are no related bugs of QEMU.
There are multiple emulators for RVV instructions in addition to QEMU, such as Spike~\cite{Spike}, NEMU~\cite{NEMU}, and Berberis~\cite{Berberis}.
Future work could focus on improving the security and reliability of emulators and hardware for SIMD instructions.

\section{Related Work}\label{sec:related}

In this section, we discuss closely related work in compiler fuzzing.

\textbf{Generation-based compiler fuzzing.}
Generation-based compiler fuzzing aims to detect compiler bugs by designing automatic program generators.
The generated programs should conform to a specific syntax and typically be well defined.
Csmith~\cite{Csmith} is widely regarded as the most influential program generator that aims to detect C/C++ compiler bugs by differential testing.
Complex heuristics are used in Csmith to avoid undefined behaviors, and hundreds of compiler-optimization bugs have been detected upon the launch of Csmith.
Multiple follow-on program generators have been developed after Csmith, each targeting specific features or programming languages.
Morisset et al.~\cite{pldi13-concurrency} develop a tool based on Csmith to detect concurrency bugs in C/C++ programs.
Lidbury et al.~\cite{pldi15-opencl} propose CLsmith, a program generator for OpenCL compilers.
Herklotz et al.~\cite{herklotz_verismith_fpga2020} propose Verismith to generate Verilog programs for FPGA synthesis tools.
Rustlantis~\cite{Rustlantis} and RustSmith~\cite{RustSmith} are random program generators for fuzzing Rust compilers.
YARPGen~\cite{YARPGen-oopsla20, YARPGen-pldi23} is a random program generator for fuzzing data-parallel programming languages; however, SIMD intrinsics are not supported by YARPGen.
Numerous outstanding contributions have been made to generation-based compiler fuzzing, and the above-mentioned work represents only a portion of them.

\textbf{Mutation-based compiler fuzzing.}
Another main approach to compiler fuzzing is mutation-based fuzzing, which generates programs by mutating seed programs (including real-world programs that are manually constructed and programs by existing generators). 
The most effective approaches in this type are the series of EMI (Equivalence Modulo Inputs) work~\cite{EMI, EMI2, EMI3}.
The EMI work mutates seed programs by removing or modifying dead regions or inserting code into live regions, at the same time ensuring that the mutated program retains the same semantics.
For C/C++ compiler fuzzing, GrayC~\cite{issta23-GrayC} mutates seed programs under the guidance of code coverage; Creal~\cite{pldi24-inject} mutates seed programs by injecting real-world programs; and MetaMut~\cite{ou2024mutators} mutates seed programs with the help of large language models for compiler fuzzing.
For JavaScript JIT compiler fuzzing, tools such as Fuzzilli~\cite{FUZZILLI}, FuzzJIT~\cite{FuzzJIT}, JIT-picking~\cite{JIT-Picking}, and OptFuzz~\cite{OptFuzz} implement mutation modules guided by domain-specific information.

\section{Conclusion}\label{sec:conclusion}
In this paper, we have proposed an approach of fuzzing compilers for \rvi{}.
We have implemented a fuzzer named \tool{} based on the ratified RVV intrinsic document in version 1.0.
\tool{} has addressed the following challenges of fuzzing compilers for \rvi{}: (i) achieving high intrinsic coverage, (ii) improving sequence variety, and (iii) without known undefined behaviors.
Experimental results have shown that \tool{} has achieved 11.5 times higher intrinsic coverage than the state-of-the-art fuzzer for RVV intrinsics. 
We have demonstrated the effectiveness of \tool{} by fuzzing three modern compilers for \rvi{}: GCC, LLVM, and XuanTie.
\tool{} has detected 13 previously unknown bugs and these bugs have been reported to the corresponding compiler developers.
Of these bugs, 10 have been confirmed and another 3 have been fixed by the compiler developers.
We expect \tool{} to open up a new direction for detecting potential compiler bugs
related to built-in functions, especially SIMD intrinsics.

\begin{acks}
This work was partially supported by Damo Academy (Hupan Laboratory) through Damo Academy (Hupan Laboratory) Innovative Research Program and National Natural Science Foundation of China under Grant No. 92464301. 
We would like to thank Ziyue Hua and Luyao Ren for discussing some technical details.
\end{acks}

\balance
\bibliographystyle{ACM-Reference-Format}
\bibliography{sample-base}

\end{document}